\documentclass{aa}

\newcommand{\teff}   {$T_{\rm eff}$~}
\newcommand{\teffe}   {$T_{\rm eff}$}
\newcommand{\teffs}   {$T_{\rm eff}$s~}
\newcommand{\rhk}    {log$(R'_{\rm HK})$~}
\newcommand{\feh}    {$[$Fe/H$]$~}

\usepackage{graphicx,natbib}
\usepackage{txfonts}
\bibpunct{(}{)}{;}{a}{}{,}
\usepackage{color}

\begin{document}

\title{The age-mass-metallicity-activity relation for solar-type stars: Comparisons with asteroseismology and the NGC 188 open cluster}

\subtitle{Letter to the Editor}

\author{D. Lorenzo-Oliveira\inst{1}, G. F. Porto de Mello\inst{1}, R. P. Schiavon\inst{2}}

\institute{Universidade Federal do Rio de Janeiro, Observat\'orio
do Valongo, Ladeira do Pedro Antonio 43, CEP: 20080-090 Rio de Janeiro, RJ, Brazil\\
\email{diego@astro.ufrj.br, gustavo@astro.ufrj.br} \and
Astrophysics Research Institute, Liverpool John Moores University,
Liverpool, L3 5RF, United Kingdom\\
\email{R.P.Schiavon@ljmu.ac.uk}}

\date{Received,Accepted}

\authorrunning{Lorenzo-Oliveira et al.}

\titlerunning{The age-mass-metallicity-activity relation for solar-type stars}

\abstract
{The Mount Wilson \ion{Ca}{II} index \rhk is the accepted standard
metric of calibration for the chromospheric activity versus age
relation for FGK stars. Recent results claim its inability to
discern activity levels, and thus ages, for stars older than
$\sim$2 Gyr, which would severely hamper its application to date
disk stars older than the Sun.}
{We present a new activity-age calibration of the Mt. Wilson index that
explicitly takes mass and \feh biases into account; these biases are implicit in
samples of stars selected to have precise ages, which have so far
not been appreciated.}
{We show that these selection biases tend to blur the activity-age
relation for large age ranges. We calibrate the Mt. Wilson index
for a sample of field FGK stars with precise ages, covering a wide
range of mass and \feh, augmented with data from the Pleiades,
Hyades, M67 clusters, and the Ursa Major moving group.}
{We further test the calibration with extensive new Gemini/GMOS
\rhk data of the old, solar \feh clusters, M67 and NGC 188. The
observed NGC 188 activity level is clearly lower
than M67. We correctly recover the isochronal age of both
clusters and establish the viability of deriving usable
chromospheric ages for solar-type stars up to at least $\sim$6
Gyr, where average errors are $\sim$0.14 dex provided that we explicitly account for the mass and \feh dimensions. We test our calibration against
asteroseismological ages, finding excellent correlation ($\rho$ = +0.89). We show that our calibration improves the chromospheric age determination for a wide range of ages, masses, and metallicities in comparison to previous age-activity relations.}
{}


\maketitle

\section{Introduction}

Stellar ages are fundamental parameters in our understanding of
the chemo-dynamical evolution of the Galaxy and other stellar
systems as well as exoplanetary systems. Ages, which are very difficult to gauge, are usually only estimated through methods
optimized to restricted classes of stars since these are indirect
parameters inferred from the time evolution of a range of
observational quantities that do not uniquely characterize this range. Strong spectral lines
are useful indicators of stellar chromospheric activity (CA)
that is physically linked to the efficiency of angular momentum evolution.
The stellar rotation and CA in single main-sequence stars decay
monotonically with time, under the action of the torque produced
by the magnetized stellar wind, as it is a potential indicator of age \citep{soderblometal1991}. The Mount Wilson (MW) project \citep{baliunasetal1995}
has been monitoring the widely used S index, which is the ratio of the flux in the line cores of the \ion{Ca}{II} H \& K lines and two nearby continuum regions; this S index can be converted into the \rhk index, which is defined as the absolute line excess flux (line flux - photospheric flux) normalized to the bolometic flux \citep{linsky79,noyesetal1984}. The \rhk is the standard metric in the literature to retrieve stellar ages through CA-age relations  \citep[hereafter CAR, e.g.,~][hereafter
MH08]{mamajekhillenbrand2008}.

Recent claims that the evolution of CA fluxes cannot be traced
beyond $\sim$2 Gyr \citep{pace2013}
imply that the derivation of CA ages is severely hampered for most
of the age dispersion of the Galactic disk, thereby negating its
usefulness as a tool to investigate Galactic evolution. Here we
analyze the presence, in the CA-age relation, of mass and
metallicity ($\mathrm{[Fe/H]}$) biases that are implicit in conventional
methods of selecting solar-type stars with precise isochronal
ages. We show that these biases have masked structural complexity
in the CAR and present a new calibration explicitly relating age,
activity, mass, and $\mathrm{[Fe/H]}$. We test this calibration against
asteroseismological ages and new Gemini data on the \rhk indexes
of the M67 and NGC 188 clusters (ages 4.0 and 6.0 Gyr,
respectively). As an extension of MH08 CAR, we provide the activity distribution expected for 6 Gyr solar metallicity stars anchored on 49 NGC 188 members (16$\times$ the M08 sample for NGC 188), constraining the activity average and dispersion beyond the solar age. 

\section{Biases in the age-chromospheric activity relation}

The authors of MH08 pointed out a slight positive trend of CA with color index ($B-V$) in the sense that hotter (more massive) stars appear less active than cooler ones; yet it was not clear that this trend extended beyond the Hyades age. Also, it was proposed that the M67 cluster presented the opposite trend of that observed in younger clusters, suggesting that evolutionary effects on color indices might affect the gauging of CA levels. Indeed, as solar-type stars age away from the Zero Age Main Sequence (ZAMS), \teff swings in the Hertzsprung-Russell (HR) diagram are bound to introduce biases into the flux calibration of the CAR. The inclusion by MH08 of an explicit color correction in their CAR hints at the presence of such degeneracy,
possibly making standard age-activity calibrations not uniquely adequate for
different levels CA.

However, even though mass and/or color terms may be explicitly
incorporated into the CAR, a deeper reason for these corrections
lies in their effects in the convective efficiency, which
establishes the theoretical connection between chromospheric and
rotational evolution and is usually represented by the Rossby
number \citep{barneskim2010}. \cite{noyesetal1984} found a strong
correlation between the Rossby number (defined as the ratio of the
rotational period and the convective overturn time $\tau_{\rm C}$)
and \rhk in FGK stars. Empirically, metallicity is expected to affect
$\tau_{\rm C}$, since  stars that are more metal rich have deeper convection
zones and thus longer $\tau_{\rm C}$, other parameters being
equal. This structural connection was already hinted at by
\cite{lyraportodemello2005} in their analysis of H$\alpha$ fluxes
and ages. However, such structural effects cannot be separated
for the case of the \ion{Ca}{II} lines from the intrinsic
observed profile which, unlike H$\alpha$ \citep{fuhrmannetal1993},
responds directly to metallicity. Thus metal-poor stars have shallower
\ion{Ca}{II} profiles that mimic high levels of chromospheric
fill-in, and thus appear more active and younger than metal-rich
stars at a given \teffe, or mass \citep{rochapintomaciel1998}.

The selection of stars with precise ages automatically packs these
biases into stellar samples. \cite{pace2013} (hereafter P13)
selected stars with small age uncertainties ($<$ 2 Gyr or
$\sigma_{\rm age}$/age $<$ 30\%), from the Geneva-Copenhagen
survey (GCS) of photometric atmospheric parameters
\citep{casagrandeetal2011}, in addition to data from clusters spanning from
0.5 to 6 Gyr. Pace crossed this sample with \rhk indices and
reported the absence of CA evolution beyond $\sim$2 Gyr. The
difficulty in building a sample of stars with precise ages is not
only dependent on the quality of the atmospheric parameters.
Demanding precise ages implies that most stars in such a sample
are more massive than the Sun because of the need for a sizable
detachment from the ZAMS to enable a reliable age determination.
Besides, such a sample necessarily has a large metallicity range:
young stars are preferably more metal rich because of the age-metallicity
relation, and possess deeper H \& K profiles, thereby mimicking a more
subdued CA. Also, in order for young stars to make their way into
a sample of objects with precise ages they tend to be farther
from the ZAMS and thus more massive. A higher mass decreases their
convective efficiency (other parameters being equal) and they appear less active. These two effects combine to lower their CA
level as deduced from \rhk. Conversely, older stars are mostly
less massive as a result of age selection effects and appear more active
owing to a heightened convective efficiency. They also tend to
be more metal poor, and their shallower H \& K profiles mimic a
higher level of CA. The net result is to strongly blur the range
of \rhk between young and old stars, thus masking the intrinsic
structure in the age-CA plane.

\section{The age-mass-metallicity-activity relation}

We present a new version of the CAR by explicitly considering its
mass and metallicity dimensions. Our sample is composed of field stars
selected from \cite{adibekyanetal2012}; in addition to field stars, we further selected objects from the Pleiades (0.1 Gyr) and Hyades (0.6 Gyr)
clusters and the Ursa Major (0.3 Gyr) moving group from
\cite{lorenzooliveiraetal2016}. Field stars
were required to have age errors $\leq$ 25\%. From
\cite{giampapaetal2006} (hereafter G06) we took data for the 4.0 Gyr (see sec. 3) M67 cluster. The sample totals 222 stars spanning the 0.75 $<$
M/M$_\odot$ $<$ 1.40 and $-$0.75 $<$ \feh $<$ $+$0.45 ranges. The
\rhk values come from a variety of sources, where the main sources are
\cite{arriagada2011}, \cite{duncanetal1991},
\cite{grayetal2003,grayetal2006}, \cite{henryetal1996},
\cite{isaacsonfischer2010}, \cite{jenkinsetal2011},
\cite{schroederetal2011}, and \cite{wrightetal2004}. Ages and masses were
inferred from the Yale isochrones \citep{kimetal2002}. 

\begin{figure}
\centering
\includegraphics[width = 0.40\textwidth]{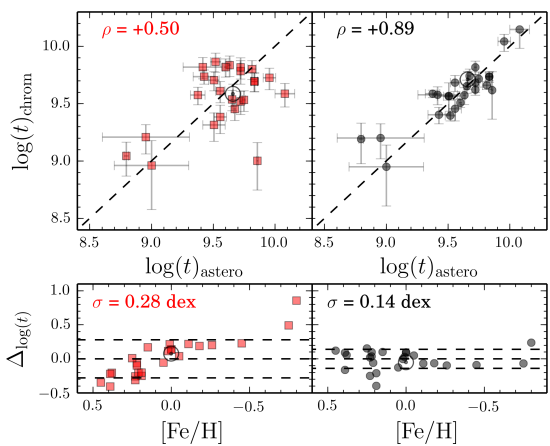}
\caption{\textit{Upper panels}: Comparison of asteroseismological with chromospheric ages
from our calibration (black circles) and MH08 (red squares) for a sample of stars spanning a wide
interval in age, mass, and metallicity. The dashed line is the 1:1 relation. \textit{Lower panels}: The difference between the age estimates $\Delta_{\log(t)}$ ($\log(t_{astero})$ - $\log(t_{chrom})$) as a function of \feh. The dashed lines stand for $\pm$ 1$\sigma$ chromospheric age uncertainty.} \label{seismo-chrom}
\end{figure}

The age-mass-metallicity-age relation (hereafter AMMAR) was
derived from an iterated, re-weighted least-squares regression,
yielding
\begin{equation}\label{eq:mod_crom_comp}
\begin{split}
\log(t) = \beta_0 + \beta_1\log(R^\prime_{HK}) + \beta_2[Fe/H] + \beta_3\log(M/M_\odot) + \\ + \beta_4\log(R^\prime_{HK})^2,
\end{split}
\end{equation}
where $\beta_0$ = $-$56.01 $\pm$ 4.74; $\beta_1$ = $-$25.81 $\pm$ 1.97; $\beta_2$ = $-$0.44 $\pm$ 0.06;  $\beta_3$ = $-$1.26 $\pm$ 0.25; and $\beta_4$ = $-$2.53 $\pm$ 0.21. The standard deviation of the relation is $\sim$0.14 dex in
log(age in years). Activity is the most significant variable in the
regression (>12$\sigma$) but mass and metallicity terms also show a strong statistical significance with 5$\sigma$ and 8$\sigma$, respectively $-$ a crucial
feature for the retrieval of reliable ages for old and/or evolved
stars. As stars evolve to lower levels of $\log(R^\prime_{HK})$, these terms
increase in their importance to explain the variance of the age
variable. We have compared our chromospheric ages
(eq.\ref{eq:mod_crom_comp}) and those derived from the age-color-activity relation (hereafter ACAR) from MH08 to asteroseismological ages in a
sample of 26 stars spanning 0.80 $<$ M/M$_\odot$ $<$ 1.57 and
$-$0.80 $<$ \feh $<$ $+$0.46 with ages from 0.4 to over 10.0 Gyr, including the Sun
\citep[][and others]{vauclair08,tang11,mosser2008,yildiz08}. The ACAR from MH08 combines the chromospheric activity-Rossby number and girochronology relations that are anchored on $\tau_C$ \citep{noyesetal1984} and ($B-V$) color index \citep{vanleeuwen07}. In the upper diagrams of Fig~\ref{seismo-chrom} we plot the results of 10$^{\rm 4}$ Monte
Carlo simulations of ages derived from eq.\ref{eq:mod_crom_comp} (black circles) and MH08 ACAR (red squares)
by spanning 4$\sigma$ intervals around each star's mass/color, metallicity, and
\rhk values. The differences between asteroseismological and chromospheric ages, as a function of \feh, are shown in the lower panels. The agreement of the AMMAR with the
asteroseismological ages is excellent: the standard deviation is $\sigma$ $\approx$ 0.14
dex and the linear correlation coefficient is $\rho$ $\approx$ 0.89 (p-value = 10$^{-7}$ \%). The MH08 ACAR shows higher scatter, weaker correlation ($\sigma$ = 0.3 dex, $\rho$ = $+$0.5), and strong residual [Fe/H] trend in the sense that higher metallicity stars tend to have higher chromospheric ages in comparison to asteroseismology. In addition, we tested the MH08 CA-age relation against the asteroseismological ages and, in comparison to ACAR, we found a similar correlation and residual [Fe/H] trend. Using the solar \rhk = $-$4.906 (MH08), the eq.\ref{eq:mod_crom_comp} yields 5.2 Gyr in good agreement with its canonical age, within 15\%. These results reinforce the need for considering a new approach to obtain consistent chromospheric ages for a wide range of masses, metallicities and activity levels.

\begin{figure}
\centering
\includegraphics[width = 0.30\textwidth]{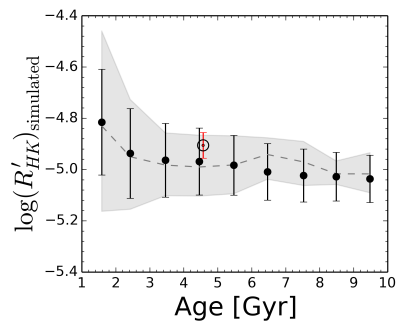}
\caption{Plot of the mean \rhk and dispersion (black dots with bars), calculated by inverting our eq. \ref{eq:mod_crom_comp} and ploted against isochronal ages, for stars selected in the GCS, under the same criteria as P13. The shaded area is the 68\% confidence interval and the dashed curve is the most probable activity value for a given age step.} \label{lshaped}
\end{figure}

A suitable test for the AMMAR is to try to reproduce the lack of an activity evolution scenario observed by P13. In this respect, we selected stars from the GCS, using their Padova mass and age estimates (16\%, 50\%, and 84\% confidence intervals) and metallicity data, following the same sample selection procedure as in P13. We kept only dwarfs and separated them
in bins of 1 Gyr. We computed their mass, age, and metallicity
distributions plus the dispersions in each age bin and
calculated the related \rhk distribution inverting
eq.\ref{eq:mod_crom_comp}, from which we obtained mean \rhk and
dispersions. In Fig~\ref{lshaped} we plot our results for the $\log(R^\prime_{HK})$-age plane and explicit ignore the metallicity and mass dimensions. We found a constant metallicity dispersion ($\approx$0.2 dex) along the age domain, which can be understood as an additional source of scatter in the age-AC diagram. In addition, the mean stellar mass representative of each age bin strongly varies with age. The sample is increasingly biased toward higher mass stars as we consider the younger stars. This feature qualitatively explains the reduced activity levels observed in each age domain in  light of AMMAR. For instance, around 1 Gyr, the typical mass is $\approx$ 1.3 M/M$_\odot$ and, after 7 Gyr, it becomes possible to select 1 M/M$_\odot$ stars. Another consequence of the selection effects is that the mean activity values overlap strongly among the age intervals and maintain large dispersions throughout the age range of the sample. Young stars show a particularly high \rhk dispersion and older stars tend to pile up in essentially the same \rhk level with a slightly lower dispersion. In comparison, we also show
the activity distribution of P13 in Fig. \ref{lshaped}. The shaded
area is the 68\% confidence interval and the dashed curve stands
for the most probable activity value for a given age step. The
features of our calculations are in good agreement with P13
results, and under these conditions no CAR can be gleaned from the
data.

\section{Chromospheric activity in old clusters: NGC 188}

Clusters are often employed as testbeds of CAR relations as they
offer a coeval stellar population with similar metallicity, allowing the
direct gauging of age (MH08) and metallicity \citep{rochapintomaciel1998}
effects. Within a given cluster population, stellar mass effects
can also be quantified. The main pitfall of introducing cluster
stars into the CAR is the very narrow metallicity interval of accessible
clusters, negating the metallicity dimension of the relation in large age
intervals. Systematic observations aimed at deriving the CA level
of open clusters are very scarce in the literature and no study
has ever tackled clusters with both non-solar metallicity and age beyond solar; the oldest cluster with a dedicated CA analysis is the 4
Gyr old, solar metallicity cluster M67 (G06).

NGC 188 is a solar-metallicity, 6 Gyr old cluster that offers a prime
opportunity to test the feasibility of extracting CA ages, by
means of the AMMAR, for stars substantially older than the Sun. We
acquired Gemini-N/GMOS spectra at R = 2,500 resolution of the M67
and NGC 188 clusters in a series of 2014$-$2015 observing runs. 
We employed the B1200 grating and 0.75'' slit width in a total of 9.7 observed hours:
the mean S/N  in the continuum was in excess of 60 for all
targets. Membership and color data come from
\cite{yadavetal2008}, \cite{gelleretal2008}, and
\cite{gelleretal2015}. We derived \teff from the relations of
\cite{casagrandeetal2011} (typical errors are $\leq$ 150 K for all
stars, including color, \teff calibration, and metallicity errors), and
computed luminosities \citep[bolometric corrections
from][]{flower1996}. Distance moduli and color excess come from
\cite{yadavetal2008} and \cite{meibom09}. Isochronal ages
derived \citep[][ the same used to derive
eq.\ref{eq:mod_crom_comp}]{kimetal2002} for the
cluster HR diagrams are 4.0 $\pm$ 0.5 and 6.0 $\pm$
0.5 Gyr for M67 and NGC 188, respectively. These age estimates are in agreement with \cite{yadavetal2008} (M67, 4.15 $\pm$ 0.65 Gyr) and \cite{meibom09} (NGC 188, 6.2 $\pm$ 0.2 Gyr).

\begin{figure}
\centering
  \begin{minipage}[t]{0.49 \linewidth}
    \resizebox{\hsize}{!}{\includegraphics{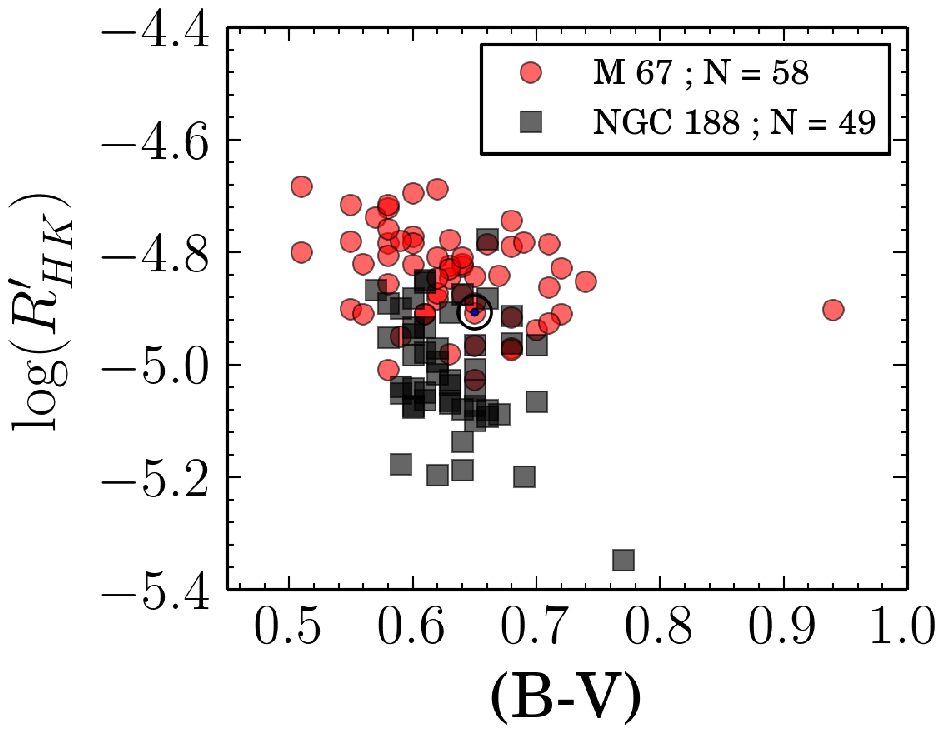}}
    
  \end{minipage}
  \begin{minipage}[t]{0.49 \linewidth}
    \resizebox{\hsize}{!}{\includegraphics{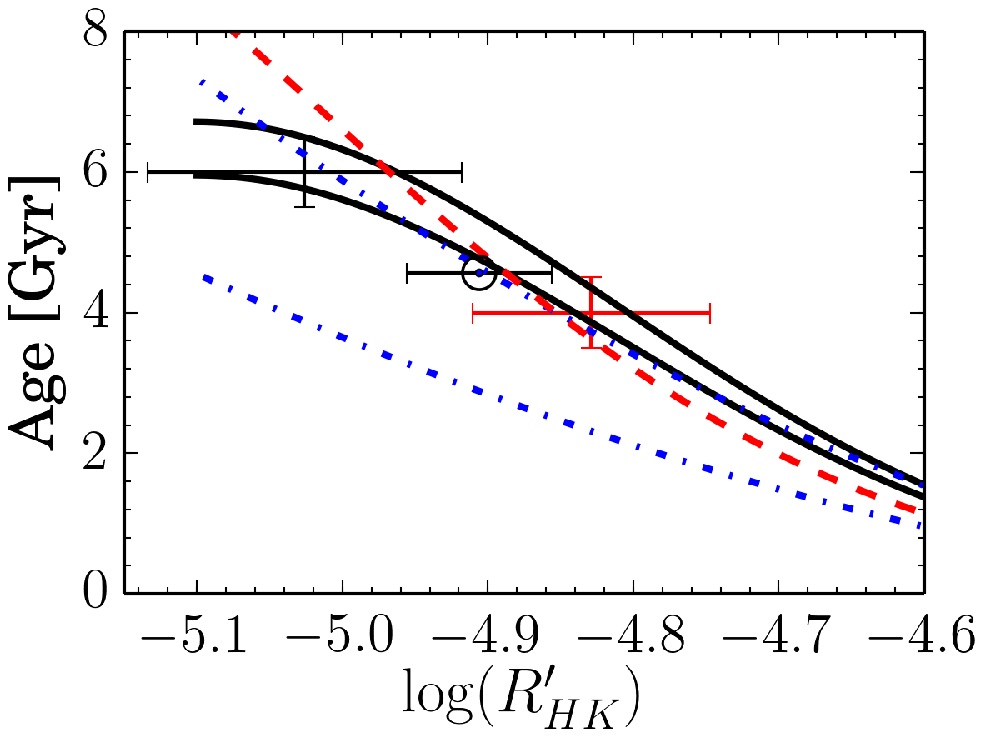}}
    
  \end{minipage}
  \caption{\textit{Left:} Observed distribution of \rhk and (B-V) in the M67 and NGC 188 clusters. The number of stars in each cluster's average is
indicated. The distribution of colors, and thus also of \teffs and
mass, is very similar in the two clusters. The Sun is plotted as
the $\odot$ symbol. \textit{Right:} \rhk for M67, NGC 188, and the Sun plotted against the age-mass-metallicity-activity relation for 1.0
(upper) and 1.1 solar masses (lower solid line). The blue dash-dotted lines stand for MH08 ACAR for ($B-V$) = 0.6 (lower) and 0.7 (upper). The red dashed line is the CA-age relation from MH08. Vertical bars are age uncertainties; horizontal bars are observed dispersions in log$(R'_{\rm HK})$.}
 
  \label{age-calibration}
\end{figure}

We applied blanketing corrections \citep{halllockwood1995} to the
spectra and converted the instrumental HK indexes to the system of
G06 by means of 12 common stars. The HK index conversion error is 15 m\AA\, ($\leq$ 10\%), which is compatible with the variability expected for M67 (G06). The HK indexes were converted into \rhk using 76 M67 stars observed by G06 and MH08. The HK-$\log(R^\prime_{HK})$ transformation was performed as a function of $(B-V)$ and  the error was 0.01 dex. Then, our M67 Gemini sample was
augmented by G06 members. The distribution of
the \rhk for the two clusters is shown in
Fig~\ref{age-calibration} (left panel); uncertainties are estimated
considering errors in photon counting, normalization, line
blanketing, colors, and transformation equations, amounting to
$\leq$ 0.10 dex for all stars. The $(B-V)$ distribution in the
two clusters is very similar, and we are thus comparing stars with
similar \teffe, minimizing any color effect that is prone to distorting the
\rhk distributions.

The median \rhk and dispersions for M67 and NGC 188
are $-$4.83 $\pm$ 0.08 and $-$5.03 $\pm$ 0.10,  respectively. Our mean figure for the
\rhk of NGC 188 agrees well with \rhk = $-$5.08 from
\cite{soderblometal1991} and differs from the \rhk = $-$4.70 used by
P13 (see his Fig. 2) based on only three stars that are slightly hotter than the 49 stars of our sample. We confirm, using the largest sample of NGC
188 stars ever, a significantly lower level of CA with respect to M67.
The probability (Anderson-Darling test) that the two distributions
of Fig~\ref{age-calibration} (left panel) are statistically indistinguishable
is 0.03 \%. In Fig~\ref{age-calibration} (right panel) we plot the \rhk values
of the Sun, M67 and NGC 188 over an extrapolation of eq. \ref{eq:mod_crom_comp} to
\rhk $<$ $-$5.1 (for solar metallicity) for two distinct masses around
the solar value. We emphasize that the NGC 188 data were not used to derive the eq.\ref{eq:mod_crom_comp}, so  it is an interesting oportunity to check the consistency of our approach for stars older than the Sun. The
chromospheric ages from eq.\ref{eq:mod_crom_comp} for M67 and NGC 188 are 4.1 $\pm$ 1.2 and 5.4 $\pm$ 0.8 Gyr, respectively, in good
agreement with the canonical isochronal ages. The ACAR from MH08 gives slightly lower age estimates of 3.0 $\pm$ 1.2 Gyr (M67) and 4.6 $\pm$ 1.9 Gyr (NGC 188) in agreement with our results within the uncertainties. The reason for these lower age estimates using ACAR is possibly due to the lack of old and inactive calibrating stars in rotation-activity and girochronology relations of MH08. In comparison to ACAR, M67 and NGC 188 chromospheric ages derived through MH08 age-activity relation are in better agreement with the canonical isochronal ages of 3.6 $\pm$ 1.3 and 7.1 $\pm$ 1.8 Gyr, respectively. However, it should be stressed that, unlike the ACAR, the CA-age relation from MH08 were calibrated using 76 M67 members from G06 and 3 NGC 188 members from \cite{soderblometal1991}. Therefore, our approach can be considered as an extension of MH08 calibration for stars with different masses and metallicities. Our results are also in
line with those of \cite{barnesetal2016}, who find a well-defined
progression in the stellar rotational periods of the clusters NGC
6811, NGC 6819 \citep{meibom11,meibom15}, and M67, up to the rotational period of the Sun,
thereby supporting age determinations linked to CA for stars at least as old as the Sun. Their gyrochoronological age for M67
is 4.2 $\pm$ 0.7 Gyr, which is in excellent agreement with our
chromospheric value.

We stress that our results do not invalidate the analysis of P13:
observational biases, the multiparametric nature of the
problem, and perhaps the age-metallicity relation conspire to produce a
complex behavior of the age-activity relation. Even though our
age-mass-metallicity-activity relation is in line with independent
observational constraints. This should not be taken as final proof
that the evolution of CA can be traced for old stars. Further
inquiry into the suitability of the \rhk index in wider, well-populated domains of mass, metallicity, age, and evolutionary states are
clearly necessary.

\section{Summary and conclusions}

We show that mass and metallicity biases are inevitably present in
samples of stars selected to have small errors in age, severely
distorting the intrinsic distribution of the chromospheric
activity $versus$ age plane. The result is a dilution of the decay
of chromospheric activity (CA) with time, apparently hampering the
derivation of ages through CA beyond $\sim$2 Gyr. These biases can
be corrected by means of an age-mass-metallicity-CA relation, which
successfully reproduces stellar asteroseismological ages up to 10
Gyr. We further test this calibration by measuring the \rhk index
of the 6.0 Gyr-old NGC 188 cluster from new and extensive
Gemini/GMOS data. The CA level of NGC 188 is clearly lower than in
the well-studied, 4.0 Gyr-old cluster M67. Our calibration
successfully recovers correct ages for both clusters:
chromospheric ages can be derived within $\sim$0.15 dex. We show that a more complete approach of the CA-age
relation, including more variables, appears promising and may
imply the ability of the age-mass-metallicity-activity relation to
recover reliable stellar ages well beyond the solar age. Present
calibrating samples, however, are far from representative of
the whole relevant domain of mass, age, metallicity, and stellar activity.
Particularly, further data on open clusters with a wide range of
age and metallicity are essential to test this approach and possibly push
the feasibility of chromospheric age determinations to the full
range of age and metallicity of the Galactic disk.


\begin{acknowledgements}

We would like to acknowledge the anonymous referee, whose comments have unquestionably led to an improved paper. DLO and GFPM acknowledge grants and scholarships from CNPq and CAPES. We thank the support of Gemini Observatory, operated by the
Association of Universities for Research in Astronomy, Inc., on
behalf of the international partnership of Argentina, Brazil,
Canada, Chile, and the USA.

\end{acknowledgements}

\end{document}